# BOOLEAN LOGIC WITH FAULT TOLERANT CODING

B. Baykant ALAGÖZ[*]

**Abstract:** *Error detectable and error correctable coding in Hamming space was researched to discover possible fault tolerant coding constellations, which can implement Boolean logic with fault tolerant property. Basic logic operators of the Boolean algebra were developed to apply fault tolerant coding in the logic circuits. It was shown that application of three-bit fault tolerant codes have provided the digital system skill of auto-recovery without need for designing additional-fault tolerance mechanisms.*

**Keyword:** Fault tolerance, error correction, fault masking

## 1. Introduction:

There have been several methods developed to implement fault tolerant logic. Most popular methods was Triple Modular Redundancy (TMR), parity check, read-solomen encoder ….etc. These methods were mostly based on designing additional circuit that was responsible to performing fault masking or error correction. Although, they can correct error resulting from fault of protected blocks, these circuits themselves were vulnerable to faults on them. It raises the question of what if fault masking circuit was faulty? In order to deal with this major problem of fault masking, we suggest technique referred as fault tolerant coding instead of additional fault-masking circuitry. Thus, fault masking would not be achieved by an additional fault-masking circuitry; instead, it will be done in level of coding by proposed fault tolerant coding technique. This technique deals with raised errors at the level of coding, instead of level of functionality (circuatry).

Conventional logic system applies Boolean equations in one bit Hamming space [1] that has two level apart from one bit distance. Unfortunately, one bit hamming coding isn't capable of detecting or correcting errors. Because there isn't any other level that can be reserved for error detection and error correction in the one bit hamming space. Additional circuits, which are solely deal with detection and correction of errors, are used for having fault tolerance feature.

In this paper, we develop a logic design methodology implementing fault tolerance in coding level. For this proposes, Boolean algebra implementation on 3 bit hamming space was researched to benefit from its one bit distance error correction capacity for the fault tolerant coding.[1] 3 bit hamming space has eight level. Two levels of them will be used for two node of Boolean lattice and the rest level will reserved for error detection and correction proposes. Basic operations of Boolean algebra, which are and (intersection), or (Union), not (Complementation) [2], were designed corresponding to three bit fault tolerant coding space. Those fault tolerant operators using three bit coding to implement Boolean Algebra was called T_and, T_or and T_not operators in order to distinguish regular and, or and not operators using conventional one-bit coding.  Relevant logic gates to perform T_and, T_or and T_not operation was designed by conventional logic gates using one-bit coding. Usage of one bit coding operators to implement operator working for higher bit coding simplifies applying higher bit coding to today's logic design methods. In Figure 1, levels of conventional logic system using one-bit coding and levels of fault tolerant logic system using $n$-bit coding was illustrated to better understanding. Level of fault tolerant operators working on $n$-bit hamming space accommodates on the conventional operators working on one-bit hamming space. According this scheme, one-bit coding expands to $n$-bit coding before implementing logic function.

---

[*] B. Baykant Alagoz  (alagozb@oncubilim.net , alagozb@yahoo.com)

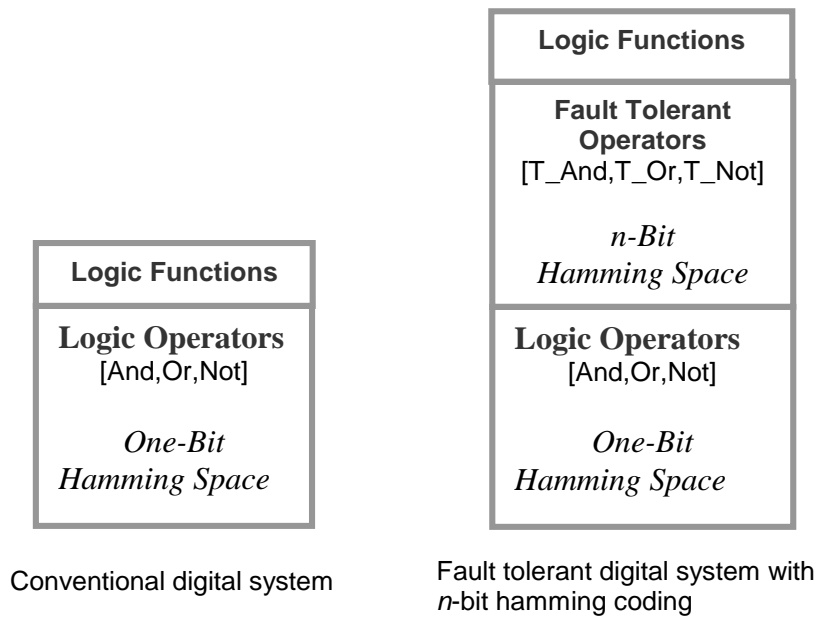

**Figure 1.** Architecture of Fault Tolerant Logic and Conventional logic

## 2. An overview of Fault Tolerant Coding In One-Two-Three Bits Hamming Spaces:

a) Basics Of Fault Tolerant Coding And Error Correction Philosophy:

Let consider $n$-bit coding of Boolean lattice, there has been $2^n$ number of level and accordingly code word. Two out of $2^n$ code words are reserved for coding of Boolean lattice highest and lowest nodes and they were called pole codes (Pole Code_0 and Pole Code_1). Pole Code_0 is the smallest element and Pole Code_1 is the largest element of lattice. There will stay $2^n - 2$ residual codes that can be utilized for error detection and correction proposes. Those residual codes were referred as faulty codes. Error detection is done by detecting any of the $2^n - 2$ faulty codes. Error correction can be done by transition of the any of $2^n - 2$ faulty codes to the nearest pole codes. As these transitions from faulty codes to pole codes is done in the direction of the pole which has minimum hamming distance to these faulty codes, considerable bit error correction can be obtained by mean of reducing faulty codes in system. This transition policy brings about attraction fields toward to pole codes in finite hamming space as represented in Figure 2. As long as faulty codes exist in space, they will fall trough the poles that is the closest in hamming space. In Figure 2, Hamming distance curvature graph illustrates transition directions (Attraction field) to pole codes.

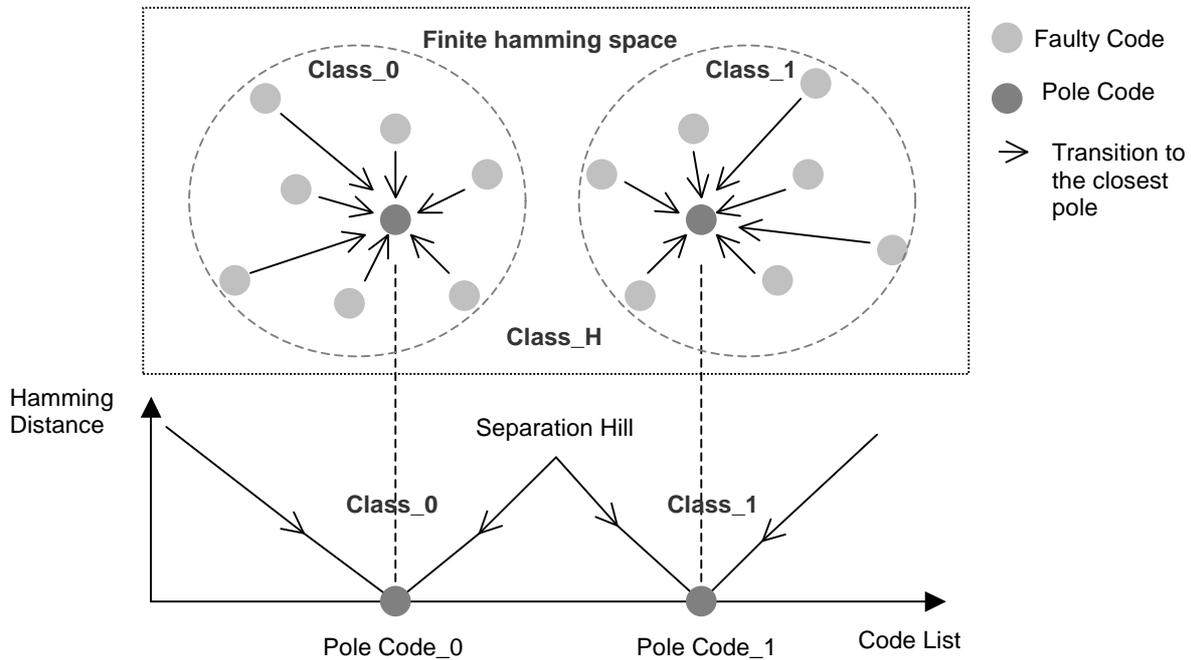

**Figure 2.** Transition mechanism for Faulty Codes

Faulty codes that has equal hamming distance to each pole code are not correctable by tolerant coding. This case coincides faulty code place onto separation hill as represented in Figure 2. Collection of faulty codes transiting to Pole Code_0 and Pole Code_0 itself compose Class_0 group and collection of faulty code transiting to Pole Code_1 and Pole Code_0 itself compose Class_1 group. Separation hill constitutes a boundary between Class_0 and Class_1. All the codes in a Class_0 or Class_1 group are correctable and detectable. If it is on the separation hill, the code is only detectable but not correctable. These codes placing on the separation hills are grouped in Class_H. Whenever a correctable code that belongs to Class_0 or Class_1 exists in hamming space, it will be turned into a pole code at next processing of Boolean algebra operation. By this transition mechanism, fault masking could be managed in coding level instead of an additional correction circuitry.

For an $n$-bit coding Boolean lattice, number of different code selection for two pole code can be given by permutation formula as following,

$$_{2^n}P_2 = \frac{2^n!}{(2^n - 2)!} \qquad (1)$$

By considering hamming space, each different pole selection forms a new transition field in the code space according to hamming distance to selected pole codes. Different pole selection in hamming space may result to have different faulty tolerance properties. In this paper, pole code selections in $n$-bit hamming space are represented in format of $(PoleCode\_0, PoleCode\_1)_n$. For example, $(0,7)_3$ refers a fault tolerant coding of which, Pole Code_0 is 0 and Pole Code_1 is 7 in 3 bit hamming space. When pole codes were selected such a way that each pole code are the most distant each other according to hamming distance, it has code constellation in which most of faulty code will position between poles. In next section, 1-bit, 2 bits and 3 bits hamming space are going to be evaluated respect to coding properties, error detection and correction capabilities in enlightenment of following statements,

- A faulty code is correctable and detectable, if it is found in either Class_0 or Class_1.
- A faulty codes are only detectable, if it is found in Class_H.
- Pole code selections vary member codes in Class_0, Class_1, Class_H sets.

b) Coding in One-Bit Hamming Space:

One-bit hamming space has two codes and all possible codes in the space are used as pole codes. In the Table 1, possible coding in one-bit hamming space were listed.

**Table 1.** Possible Coding In One-Bit Hamming Space

| Pole Code_0 | Pole Code_1 | Class_0 | Class_1 | Class_H | Distance Between Poles |
|---|---|---|---|---|---|
| 0 | 1 | 0 | 1 | - | 1 |
| 1 | 0 | 1 | 0 | - | 1 |

Basic properties of one bit coding,

- It can code two levels. All possible code words are $\{0,1\}$. These code words are reserved for representation of two nodes of Boolean lattice. Two different coding can be done for Boolean lattice. All of them is listed in Table 1.
- There isn't any faulty code that can be used for error detection and correction purposes. So, Class_0 and Class_1 do not contain any faulty code.

Coding used in conventional logic design is $(0,1)_1$ coding and it doesn't have any faulty codes, which is correctable or detectable. Therefore, an additional logic circuit is required to perform error detection or error correction.

c) Coding in Two-Bit Hamming Space:

Two-bit hamming space has four codes and it provides two faulty codes. In the Table 2., noteworthy coding in two-bit hamming space was listed. Basic properties of one bit coding;

- It can code four levels. All possible code words are $\{0,1,2,3\}$. Depending on selection of pole codes, two type-coding properties was observed; first type is distant selection in which poles are selected with maximum distance to each other. This type coding is seen in first four rows in Table 2. Hamming distance between poles are 2. The other type is nearby selection in which poles are selected in close proximity. This type of coding is given in last four rows in Table 2. Distance between these poles are 1 bit.
- There are two faulty codes that can be used for error detection and correction purposes. In the distant coding type, Class_0 and Class_1 do not contain any faulty code. Both of faulty codes reside on Class_H group. This is why; they aren't correctable, but detectable. On the other hand, in the nearby coding type, Class_0 and Class_1 groups contain faulty codes. Therefore; faulty codes of nearby coding are both detectable and correctable.
- From application point of view, nearby coding style for two-bit hamming space has advantage of bearing correctable faulty codes.

**Table 2.** Some noteworthy Coding from 12 Possible Coding in Two-Bit Hamming Space

| Pole Code_0 | Pole Code_1 | Class_0 | Class_1 | Class_H | Distance Between Poles |
|---|---|---|---|---|---|
| 0 | 3 | 0 | 3 | 1,2 | 2 |
| 3 | 0 | 3 | 0 | 1,2 | 2 |
| 1 | 2 | 1 | 2 | 0,3 | 2 |
| 2 | 1 | 2 | 1 | 0,3 | 2 |
| 1 | 3 | 1,0 | 3,2 | - | 1 |
| 3 | 1 | 3,2 | 1,0 | - | 1 |
| 0 | 2 | 0,1 | 2,3 | - | 1 |
| 2 | 0 | 2,3 | 0,1 | - | 1 |

Transition graph for a nearby coding example of two-bit hamming space is given in Figure 3.

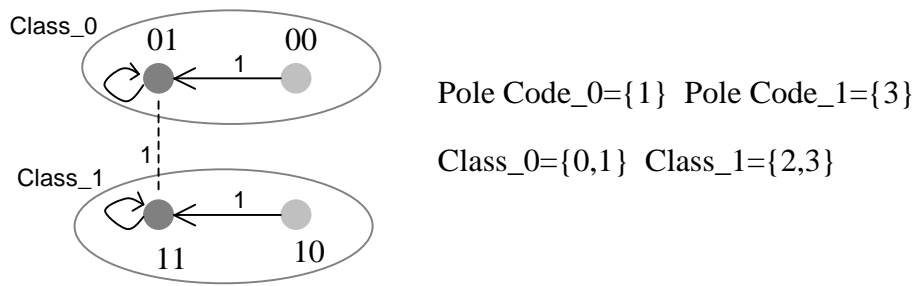

Pole Code_0={1}  Pole Code_1={3}

Class_0={0,1}  Class_1={2,3}

**Figure 3.** Transition graph for a nearby coding of the two-bit hamming space ($(1,3)_2$ tolerant coding)

According the transition graph seen in Figure 3, all faulty code in one bit proximity to poles can be correctable but unfortunately pole codes are in one bit distance each other.

d) Coding in Three-Bit Hamming Space:

Three-bit hamming space has eight codes and it provides six faulty codes. In the Table 3., noteworthy coding in three-bit hamming space was listed.

**Table 3.** Some noteworthy Coding from 56 Possible Coding in Three-Bit Hamming Space

| Pole Code_0 | Pole Code_1 | Class_0 | Class_1 | Class_H | Distance Between Poles | Maximum Correctable Faulty Code Distance |
|---|---|---|---|---|---|---|
| 0 | 7 | 0,1,2,4 | 3,5,6,7 | - | 3 | 1 |
| 7 | 0 | 3,5,6,7 | 0,1,2,4 | - | 3 | 1 |
| 1 | 6 | 0,1,3,5 | 2,4,6,7 | - | 3 | 1 |
| 6 | 1 | 2,4,6,7 | 0,1,3,5 | - | 3 | 1 |
| 2 | 5 | 0,2,3,6 | 1,4,5,7 | - | 3 | 1 |
| 5 | 2 | 1,4,5,7 | 0,2,3,6 | - | 3 | 1 |
| 0 | 3 | 1,4 | 7 | 2,5,6 | 2 | 1 |
| 3 | 0 | 7 | 1,4 | 2,5,6 | 2 | 1 |
| 1 | 3 | 0,**4**,5 | 2,7,**6** | - | 1 | 2[*] |
| 3 | 1 | 2,6,7 | 0,4,5 | - | 1 | 2[*] |

[*] bold typed faulty codes in Class_0 and Class_1 are correctable code with 2 bit hamming distance to poles

Basic properties of three bit coding,
- It can code eight levels. All possible code words are $\{0,1,2,3,4,5,6,7\}$. In the 3 bit hamming space, pole code selection can be done in one, two and three bits distance. In the Table 3, some example of code selection with various distance were given to observe some properties of being correctable and detectable.
- Three bit hamming space gives us the opportunity of coding with the most distant pole selection with fully correctable faulty codes set. Coding seen in first six row of Table 3 shows such coding that pole code resides in the most distant each other and all faulty codes are in either Class_0 or Class_1. Three bit coding with 3 bit hamming distance selection provides better fault tolerance in application compared to other coding of one, two bits. $(0,3)_3$ and $(3,0)_3$ coding has some codes in Class_H. These codes are detectable but not correctable. In the $(1,3)_3$ and $(3,1)_3$ coding, although distance between pole code reduced to one bit, faulty codes 4 and 6 are correctable code with two bit distance to poles. It shows that correction of faulty codes with two-bit distance is possible in three bit hamming space in the expense of diminishing distance of pole codes to each other.

- From application point of view, three bits distant pole code selections in three-bit hamming space have advantage of all faulty codes being correctable as well as poles is wide apart each other in the space.

Transition graph for $(2,5)_3$ tolerant coding listed in the fifth row of Table 3 was illustrated in Figure 4.

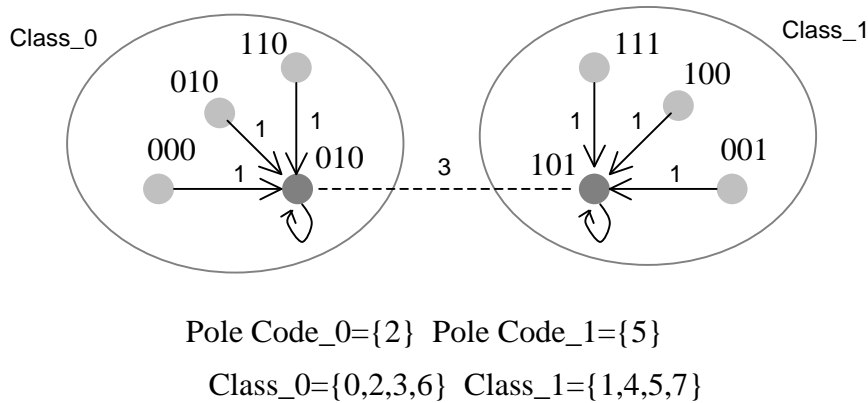

Pole Code_0={2}  Pole Code_1={5}

Class_0={0,2,3,6}  Class_1={1,4,5,7}

**Figure 4.** An example of transition graphs for a three-bits coding with three-bit distance ($(2,5)_3$ tolerant coding)

One, two and three bit hamming space were summarized for better comparison of performance of fault masking in Table 4. Three-bits hamming space exhibits the one-bit distant correction with three bit distant pole coding and digital system using this coding will correct whole faulty codes in one bit distance to pole codes and it gains tolerance against one-bit-errors (single errors). In two bits hamming space, two-bit coding doesn't guaranty correction of all one-bit errors (single errors) due to one bit distance between poles. Because, some one-bit errors would possibly coincide on the other pole rather than a faulty code and in such cases, they will not be correctable and detectable like coding in one-bit hamming space. This situation degrades fault tolerance capability of the coding. Pole code selection criteria for better fault tolerance capability can be summarized as following;
1. Pole codes should be select apart each other with maximum distance of its hamming space. This condition can be satisfied by selection of pole codes such that pole codes are complement of each other. Such as $(2,5)_3$, $(0,7)_3$ ...
2. Class_H should be empty set. All faulty code should be resided in either Class_0 or Class_1.

**Table 4.** Comparison fault masking performance of coding in hamming spaces

| Number Of Bit | Number Of Code | Number of Pole Code Selection | Number Of Faulty Codes | Largest Pole Distance | Maximum Correctable Faulty Code Distance |
|---|---|---|---|---|---|
| 1 | 2 | 2 | 0 | 1 | - |
| 2 | 4 | 12 | 2 | 2 | 1 |
| 3 | 8 | 56 | 6 | 3 | 1 |

Triple Modular Redundancy (TMR) applies $(0,7)_3$ coding. TMR has been widely used technique to mask faults in functionality level by a voting circuit.[3] Three redundant modules produce three-bit codes so that a voting circuit perform correction and encode three-bits code to one-bit codes.

## 3. Logic Design Approaches For Fault Tolerant Coding And Examples For $(2,5)_3$ Coding :

Basic idea behind the Boolean algebra implementation with fault tolerant coding is to replace pole codes (Pole Code_0 and Pole_Code_1) with the lowest and highest nodes of Boolean lattice Which are logic '0' and logic '1' symbols in conventional digital design. In every logic operation, faulty codes were treated as Pole Code_0 or Pole Code_1 according the transition graph of coding. To better express, all codes in Class_0 group are taken account as Pole Code_0 and all codes in Class_1 are taken account as Pole Code_1. This treatment of codes in Class_0 and Class_1 group naturally constitutes mechanism of faulty code transition to pole code in coding level; thus, we say that it manages correction of faulty codes in coding level. Faulty codes in Class_H group are freely accounted as either Pole Code_0 or Pole_Code_1 for the reduction of logic equations. In the Table 5 and 6, truth table of the conventional Boolean operator and truth table of the fault tolerant coding were given.

**Table 5.** Truth table of basic logic operation of Boolean algebra for $(0,1)_1$ conventional coding

| $a$ | $b$ | $a+b$ | $a.b$ | $\bar{a}$ |
|---|---|---|---|---|
| 0 | 0 | 0 | 0 | 1 |
| 0 | 1 | 1 | 0 | 1 |
| 1 | 0 | 1 | 0 | 0 |
| 1 | 1 | 1 | 1 | 0 |

**Table 6.** Truth table of basic logic operation of Boolean algebra for fault tolerant coding

| $a$ | $b$ | $a \oplus b$ | $a \otimes b$ | $\bar{a}$ |
|---|---|---|---|---|
| Class_0 | Class_0 | Pole Code_0 | Pole Code_0 | Pole Code_1 |
| Class_0 | Class_1 | Pole Code_1 | Pole Code_0 | Pole Code_1 |
| Class_1 | Class_0 | Pole Code_1 | Pole Code_0 | Pole Code_0 |
| Class_1 | Class_1 | Pole Code_1 | Pole Code_1 | Pole Code_0 |

In the Table 6, $a \oplus b$ operation is tolerant *Or* operation (T_or), $a \otimes b$ operation is tolerant *and* operation (T_and) and $\bar{a}$ operation is tolerant complementation (T_not). As seen in table, these operators take its input from Class_0 and Class_1 group and they yield pole codes depending on logic operation. This input-output relation led transitions from faulty codes to pole codes take place in every operation interval.

By considering Table 6, Pole equations respect to Pole Code_0 can be written as following;

$$(a \oplus b)_{PoleCode\_0} = Class\_0 \times Class\_0 \quad (2)$$
$$(a \otimes b)_{PoleCode\_0} = Class\_0 \times Class\_0 \cup Class\_0 \times Class\_1 \cup Class\_1 \times Class\_0 \quad (3)$$
$$(\overline{a})_{PoleCode\_0} = Class\_1 \quad (4)$$

Here, $\times$ Cartesian product operator and $\cup$ set unification operation. $(.)_{PoleCode\_X}$ representation express that equation is written corresponding to *Pole Code_X*. Equations corresponding to Pole Code_1 are given as following,

$$(a \oplus b)_{PoleCode\_1} = Class\_0 \times Class\_1 \cup Class\_1 \times Class\_0 \cup Class\_1 \times Class\_1 \quad (5)$$
$$(a \otimes b)_{PoleCode\_1} = Class\_1 \times Class\_1 \quad (6)$$
$$(\overline{a})_{PoleCode\_1} = Class\_0 \quad (7)$$

In the next section, tolerant logic operator for three bit tolerant coding was designed using conventional logic gates. For this proposes, input couple $(a,b)$ and output $f$ is expanded to three bit binary format. Output $f$ takes value of Pole code_0 or Pole code_1 in binary format in case of fault-free. Truth table using these binary inputs $a,b$ and output $f$ can be construct by guidance of equation 2,3,4,5,6 and 7. In these way, as presented in Figure 1, fault tolerant coding in three bit hamming space can be build on today's conventional logic methodology.

a) Fault Tolerant Logic Design By Component Substitution In Schematic of Conventional Logic:

Lets design T_or, T_and and T_not operators for three bit tolerant coding. Firstly, logic operator for $(2,5)_3$ tolerant coding seen in fifth row of Table 3 and Figure 4 were designed to discover basics of $(2,5)_3$ given in step 1. Then follow the steps;

**Step 1:** State $(2,5)_3$ tolerant coding structures; $PoleCode\_0 = 2$, $PoleCode\_1 = 5$, $Class\_0 = \{0,2,3,6\}$ and $Class\_1 = \{1,4,5,7\}$

**Step 2:** $T\_Or = (a \oplus b)_{PoleCode\_0}, T\_And = (a \otimes b)_{PoleCode\_1}$ and $T\_Not = (\overline{a})_{PoleCode\_1}$ sets are obtained by using equation 2, 6 and 7. ($T\_Or = (a \oplus b)_{PoleCode\_0}$ was selected instead of $T\_Or = (a \oplus b)_{PoleCode\_1}$ for design simplifications). These subsets of Cartesian product set of inputs result subscripted PoleCode_X value, for the rest of inputs in truth table, it results the other pole code.

$$(a \oplus b)_{PoleCode\_0} = \begin{cases} (0,0),(0,2),(0,3),(0,6),(2,0),(2,2),(2,3),(2,6),(3,0),(3,2),(3,3),(3,6), \\ (6,0),(6,2),(6,3),(6,6) \end{cases}$$

$$(a \otimes b)_{PoleCode\_1} = \begin{cases} (1,1),(1,4),(1,5),(1,7),(4,1),(4,4),(4,5),(4,7),(5,1),(5,4),(5,5),(5,7), \\ (7,1),(7,4),(7,5),(7,7) \end{cases}$$

$$(\overline{a})_{PoleCode\_1} = \{0,2,3,6\}$$

**Step 3:** Expand all elements of sets and $PoleCode\_0$ and $PoleCode\_1$ in three bit binary format and construct truth table corresponding to one-bit conventional coding as following,

| a | | | b | | | $T\_Or = (a \oplus b)_{PoleCode\_0}$ | | |
|---|---|---|---|---|---|---|---|---|
| $a_1$ | $a_2$ | $a_3$ | $b_1$ | $b_2$ | $b_3$ | $T\_Or_1$ | $T\_Or_2$ | $T\_Or_3$ |
| 0 | 0 | 0 | 0 | 0 | 0 | 0 | 1 | 0 |
| 0 | 0 | 0 | 0 | 1 | 0 | 0 | 1 | 0 |
| 0 | 0 | 0 | 0 | 1 | 1 | 0 | 1 | 0 |
| 0 | 0 | 0 | 1 | 1 | 0 | 0 | 1 | 0 |
| 0 | 1 | 0 | 0 | 0 | 0 | 0 | 1 | 0 |
| 0 | 1 | 0 | 0 | 1 | 0 | 0 | 1 | 0 |
| 0 | 1 | 0 | 0 | 1 | 1 | 0 | 1 | 0 |
| 0 | 1 | 0 | 1 | 1 | 0 | 0 | 1 | 0 |
| 0 | 1 | 1 | 0 | 0 | 0 | 0 | 1 | 0 |
| 0 | 1 | 1 | 0 | 1 | 0 | 0 | 1 | 0 |
| 0 | 1 | 1 | 0 | 1 | 1 | 0 | 1 | 0 |
| 0 | 1 | 1 | 1 | 1 | 0 | 0 | 1 | 0 |
| 1 | 1 | 0 | 0 | 0 | 0 | 0 | 1 | 0 |
| 1 | 1 | 0 | 0 | 1 | 0 | 0 | 1 | 0 |
| 1 | 1 | 0 | 0 | 1 | 1 | 0 | 1 | 0 |
| 1 | 1 | 0 | 1 | 1 | 0 | 0 | 1 | 0 |
| x | x | x | x | x | x | 1 | 0 | 1 |

| a | | | b | | | $T\_And = (a \otimes b)_{PoleCode\_1}$ | | |
|---|---|---|---|---|---|---|---|---|
| $a_1$ | $a_2$ | $a_3$ | $b_1$ | $b_2$ | $b_3$ | $T\_And_1$ | $T\_And_2$ | $T\_And_3$ |
| 0 | 0 | 1 | 0 | 0 | 1 | 1 | 0 | 1 |
| 0 | 0 | 1 | 1 | 0 | 0 | 1 | 0 | 1 |
| 0 | 0 | 1 | 1 | 0 | 1 | 1 | 0 | 1 |
| 0 | 0 | 1 | 1 | 1 | 1 | 1 | 0 | 1 |
| 1 | 0 | 0 | 0 | 0 | 1 | 1 | 0 | 1 |
| 1 | 0 | 0 | 1 | 0 | 0 | 1 | 0 | 1 |
| 1 | 0 | 0 | 1 | 0 | 1 | 1 | 0 | 1 |
| 1 | 0 | 0 | 1 | 1 | 1 | 1 | 0 | 1 |
| 1 | 0 | 1 | 0 | 0 | 1 | 1 | 0 | 1 |

| 1 | 0 | 1 | 1 | 0 | 0 | 1 | 0 | 1 |
|---|---|---|---|---|---|---|---|---|
| 1 | 0 | 1 | 1 | 0 | 1 | 1 | 0 | 1 |
| 1 | 0 | 1 | 1 | 1 | 1 | 1 | 0 | 1 |
| 1 | 1 | 1 | 0 | 0 | 1 | 1 | 0 | 1 |
| 1 | 1 | 1 | 1 | 0 | 0 | 1 | 0 | 1 |
| 1 | 1 | 1 | 1 | 0 | 1 | 1 | 0 | 1 |
| 1 | 1 | 1 | 1 | 1 | 1 | 1 | 0 | 1 |
| x | x | x | x | x | x | 0 | 1 | 0 |

| $a$ | | | $T\_Not = (\overline{a})_{PoleCode\_1}$ | | |
|---|---|---|---|---|---|
| $a_1$ | $a_2$ | $a_3$ | $T\_Not_1$ | $T\_Not_2$ | $T\_Not_3$ |
| 0 | 0 | 0 | 1 | 0 | 1 |
| 0 | 1 | 0 | 1 | 0 | 1 |
| 0 | 1 | 1 | 1 | 0 | 1 |
| 1 | 1 | 0 | 1 | 0 | 1 |
| x | x | x | 0 | 1 | 0 |

**Step 4:** Obtain corresponding binary logic function by using truth table seen in step 3. Solutions for $(2,5)_3$ coding were given bellow in sum of product (SOP) by $\sum(.)$ symbol and product of sum (POS) by $\prod(.)$ symbol.

$$T\_Or_1 = \prod(0,2,3,6,16,18,19,22,24,26,27,30,48,50,51,54) \quad (8)$$

$$T\_Or_2 = \sum(0,2,3,6,16,18,19,22,24,26,27,30,48,50,51,54) \quad (9)$$

$$T\_Or_3 = \prod(0,2,3,6,16,18,19,22,24,26,27,30,48,50,51,54) \quad (10)$$

$$T\_And_1 = \sum(9,12,13,15,33,36,37,39,41,44,45,47,57,60,61,63) \quad (11)$$

$$T\_And_2 = \prod(9,12,23,15,33,36,37,39,41,44,45,47,57,60,61,63) \quad (12)$$

$$T\_And_3 = \sum(9,12,23,15,33,36,37,39,41,44,45,47,57,60,61,63) \quad (13)$$

$$T\_Not_1 = \sum(0,2,3,6) \quad (14)$$

$$T\_Not_2 = \prod(0,2,3,6) \quad (15)$$

$$T\_Not_3 = \sum(0,2,3,6) \quad (16)$$

Schematics of the T_Or, T_And and T_Not gates implementing $(2,5)_3$ tolerant coding were illustrated in the Figure 5. An important point to be considered is that, by mean of these tolerant gates, any Boolean function using tolerant coding can be designed without following four step seen above for each logic function. It can be simply done by replacing logic gates in circuit of conventional $(0,1)_1$ coding with relevant tolerant gates. In this way, knowledge relating tolerant coding had been encapsulated in tolerant gates and it doesn't interfere functionality of logic equation.

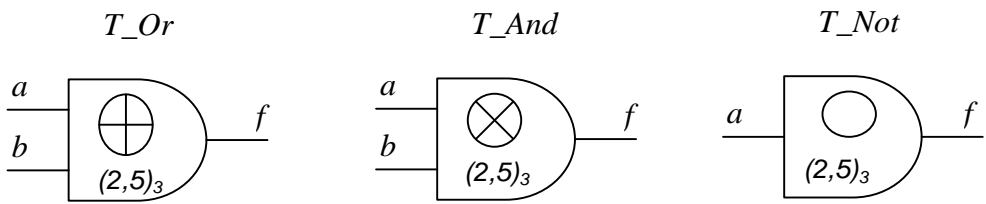

**Figure 5.** Logic gates for $(2,5)_3$ tolerant coding

Lets design XOR function for $(2,5)_3$ tolerant coding by using T_Or, T_And and T_Not gates seen in Figure 5. The design simplicity can be clearly seen by considering truth tables of XOR given in Table 7 for both one-bit conventional coding $(0,1)_1$ and $(2,5)_3$ tolerant coding. In order to facilitate design process of tolerant coding, design for tolerant coding can be done by simply replacing conventional logic gates in the schema with tolerant gates. Thus, truth tables will be analogous for each other in term of logic functionality. In the Figure 6, XOR for $(2,5)_3$ tolerant coding is designed by replacing logic gates of $(0,1)_1$ with tolerant gates of $(2,5)_3$.

**Table 7.** Truth table of XOR for one-bit coding

| Conventional XOR | | | Fault Tolerant XOR | | |
|---|---|---|---|---|---|
| a | b | F | a | b | f |
| 0 | 0 | 0 | Class_0 | Class_0 | Pole Code_0 |
| 0 | 1 | 1 | Class_0 | Class_1 | Pole Code_1 |
| 1 | 0 | 1 | Class_1 | Class_0 | Pole Code_1 |
| 1 | 1 | 0 | Class_1 | Class_1 | Pole Code_0 |

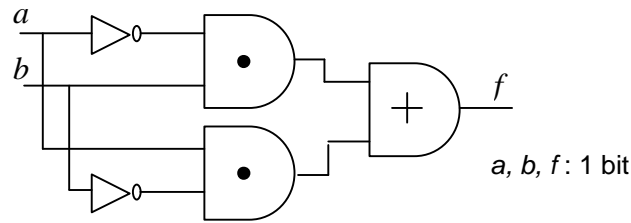

XOR function for one-bit conventional coding $(0,1)_1$

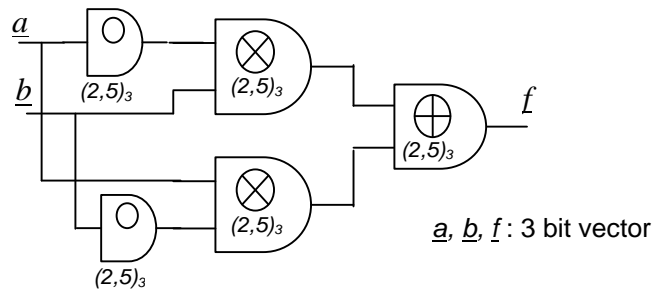

XOR function for fault tolerant coding $(2,5)_3$

**Figure 6.** XOR for $(2,5)_3$ tolerant coding by using XOR for conventional $(0,1)_1$ coding.

Sequential components such as flip-flops and more complex logic functions can be easily designed under the same perspective discussed above for $(2,5)_3$ tolerant coding. In the Figure 8, several flip-flop types designed for $(2,5)_3$ tolerant coding were illustrated.

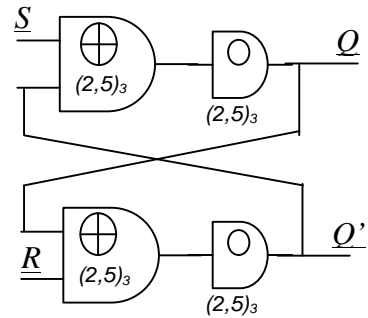

SR Flip-Flop for $(2,5)_3$ tolerant coding

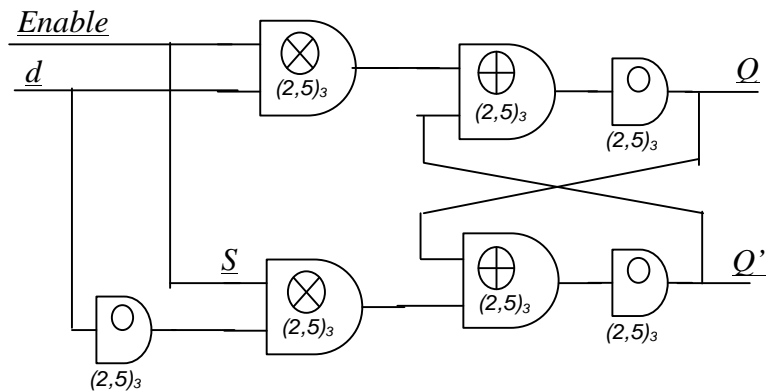

D Flip-Flop for $(2,5)_3$ tolerant coding

**Figure 7.** Some example for sequential components for $(2,5)_3$ tolerant coding

A digital IC using conventional $(0,1)_1$ coding can be easily revised to perform tolerant coding by mean of component substitution technique explained in this section. However, this method simplifies design process, it may consumes very large area of logic (resource).

b) Fault Tolerant Logic Design From Truth Tables:
In the previous section, we already discussed designing directly from Truth table for basic logic operator (*T_Or*, *T_And*, *T_Not*). In this section, we will apply it for more complex logic functions.
Lets design EXOR function directly from truth table seen in Table 7. For this proposes, write design steps discussed previous section for EXOR function.

**Step 1:** State $(2,5)_3$ tolerant coding structures; $PoleCode\_0 = 2$, $PoleCode\_1 = 5$, $Class\_0 = \{0,2,3,6\}$ and $Class\_1 = \{1,4,5,7\}$

**Step 2:** From Table 7, we can write for fault tolerant logic equation as

$$(f)_{PoleCode\_1} = Class\_0 \times Class\_1 \cup Class\_1 \times Class\_0$$

And write Cartesian product set,

$$(f)_{PoleCode\_1} = \begin{Bmatrix} (0,1),(0,4),(0,5),(0,7),(2,1),(2,4),(2,5),(2,7),(3,1),(3,4),(3,5),(3,7), \\ (6,1),(6,4),(6,5),(6,7),(1,0),(1,2),(1,3),(1,6),(4,0),(4,2),(4,3),(4,6), \\ (5,0),(5,2),(5,3),(5,6),(7,0),(7,2),(7,3),(7,6) \end{Bmatrix}$$

**Step 3:** Expand all elements of $(f)_{PoleCode\_1}$ set and $PoleCode\_1$ in three bit binary format and construct truth table corresponding to one-bit conventional coding as following,

| a | | | b | | | $(f)_{PoleCode\_1}$ | | |
|---|---|---|---|---|---|---|---|---|
| $a_1$ | $a_2$ | $a_3$ | $b_1$ | $b_2$ | $b_3$ | $T\_And_1$ | $T\_And_2$ | $T\_And_3$ |
| 0 | 0 | 0 | 0 | 0 | 1 | 1 | 0 | 1 |
| 0 | 0 | 0 | 1 | 0 | 0 | 1 | 0 | 1 |
| 0 | 0 | 0 | 1 | 0 | 1 | 1 | 0 | 1 |
| 0 | 0 | 0 | 1 | 1 | 1 | 1 | 0 | 1 |
| 0 | 1 | 0 | 0 | 0 | 1 | 1 | 0 | 1 |
| 0 | 1 | 0 | 1 | 0 | 0 | 1 | 0 | 1 |
| 0 | 1 | 0 | 1 | 0 | 1 | 1 | 0 | 1 |
| 0 | 1 | 0 | 1 | 1 | 1 | 1 | 0 | 1 |
| 0 | 1 | 1 | 0 | 0 | 1 | 1 | 0 | 1 |
| 0 | 1 | 1 | 1 | 0 | 0 | 1 | 0 | 1 |
| 0 | 1 | 1 | 1 | 0 | 1 | 1 | 0 | 1 |
| 0 | 1 | 1 | 1 | 1 | 1 | 1 | 0 | 1 |
| 1 | 1 | 0 | 0 | 0 | 1 | 1 | 0 | 1 |
| 1 | 1 | 0 | 1 | 0 | 0 | 1 | 0 | 1 |
| 1 | 1 | 0 | 1 | 0 | 1 | 1 | 0 | 1 |
| 1 | 1 | 0 | 1 | 1 | 1 | 1 | 0 | 1 |
| 0 | 0 | 1 | 0 | 0 | 0 | 1 | 0 | 1 |
| 0 | 0 | 1 | 0 | 1 | 0 | 1 | 0 | 1 |
| 0 | 0 | 1 | 0 | 1 | 1 | 1 | 0 | 1 |
| 0 | 0 | 1 | 1 | 1 | 0 | 1 | 0 | 1 |
| 1 | 0 | 0 | 0 | 0 | 0 | 1 | 0 | 1 |
| 1 | 0 | 0 | 0 | 1 | 0 | 1 | 0 | 1 |
| 1 | 0 | 0 | 0 | 1 | 1 | 1 | 0 | 1 |
| 1 | 0 | 0 | 1 | 1 | 0 | 1 | 0 | 1 |
| 1 | 0 | 1 | 0 | 0 | 0 | 1 | 0 | 1 |
| 1 | 0 | 1 | 0 | 1 | 0 | 1 | 0 | 1 |
| 1 | 0 | 1 | 0 | 1 | 1 | 1 | 0 | 1 |
| 1 | 0 | 1 | 1 | 1 | 0 | 1 | 0 | 1 |
| 1 | 1 | 1 | 0 | 0 | 1 | 1 | 0 | 1 |
| 1 | 1 | 1 | 1 | 0 | 0 | 1 | 0 | 1 |
| 1 | 1 | 1 | 1 | 0 | 1 | 1 | 0 | 1 |
| 1 | 1 | 1 | 1 | 1 | 1 | 1 | 0 | 1 |
| x | x | x | x | x | x | 0 | 1 | 0 |

**Step 4:** Obtain corresponding binary logic function from truth table seen in step 3. Solutions for $(2,5)_3$ coding were written bellow in sum of product (SOP) by $\sum(.)$ symbol and product of sum (POS) by $\prod(.)$ symbol.

$$f_1 = \sum \begin{array}{l}(1,4,5,7,8,10,11,14,17,20,21,23,25,28,29,31,32,34,35,38,40,42,43,\\ 46,49,52,53,55,57,60,61,63)\end{array} \quad (17)$$

$$f_2 = \prod \begin{array}{l}(1,4,5,7,8,10,11,14,17,20,21,23,25,28,29,31,32,34,35,38,40,42,43,\\ 46,49,52,53,55,57,60,61,63)\end{array} \quad (18)$$

$$f_3 = \sum \begin{array}{l}(1,4,5,7,8,10,11,14,17,20,21,23,25,28,29,31,32,34,35,38,40,42,43,\\ 46,49,52,53,5557,60,61,63)\end{array} \quad (19)$$

Logic reduction can be applied to reduce gate count.

d) Coding Translator Components:

Coding translator is required when more then one coding or hamming space were cooperated at the same system. They translate codes from a coding in a hamming space to any other coding in any hamming space. The coding translator takes part at the interfacing two different coding as seen Figure 11.

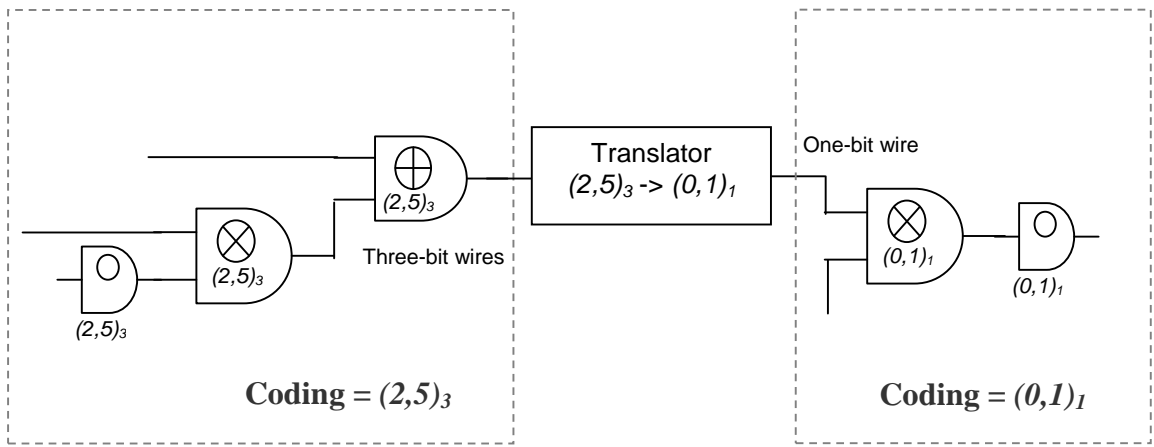

**Figure 11.** Translator interfacing between $(2,5)_3$ and $(0,1)_1$ tolerant coding

Translation from $(2,5)_3$ to $(0,1)_1$ tolerant coding can be done according to truth table seen in Table 9.

**Table 9.** Truth table of translator from $(2,5)_3$ to $(0,1)_1$ tolerant coding.

| $(2,5)_3$ Codes | | | $(0,1)_1$ Codes |
|---|---|---|---|
| a | b | c | Tr |
| 0 | 0 | 0 | 0 |
| 0 | 0 | 1 | 1 |
| 0 | 1 | 0 | 0 |
| 0 | 1 | 1 | 0 |
| 1 | 0 | 0 | 1 |
| 1 | 0 | 1 | 1 |
| 1 | 1 | 0 | 0 |
| 1 | 1 | 1 | 1 |

Logic equation in sum of product (SOP) format for translation from $(2,5)_3$ to $(0,1)_1$ tolerant coding can be written as,

$$Tr = \sum(1,4,5,7) \qquad (20)$$

Translation from $(0,1)_1$ to $(2,5)_3$ tolerant coding can be done according to truth table seen in Table 10.

**Table 10.** Truth table of translator from $(0,1)_1$ to $(2,5)_3$ tolerant coding

| $(0,1)_1$ Codes | $(2,5)_3$ Codes | | |
|---|---|---|---|
| a | $Tr_1$ | $Tr_2$ | $Tr_3$ |
| 0 | 0 | 1 | 0 |
| 1 | 1 | 0 | 1 |

Logic equation for translation from $(0,1)_1$ to $(2,5)_3$ tolerant coding can be written as,

$$Tr_1 = a \qquad (21)$$
$$Tr_2 = \bar{a} \qquad (22)$$
$$Tr_3 = a \qquad (23)$$

Translators have two basic properties;

- They do not change logic value but they translate a word from a coding to the other.
- They can exhibit correction capability.

e) Fault Tolerance Mechanism And Flaws In Applications:

Before correction of the faulty code, a faulty code must physically appear on the logic system. There are two phases in tolerance mechanism.

- First phase is the *fault appearance phase* of faulty code in which faulty codes were seen at the output of any tolerant gates.
- Second phase is the *transition phase* of faulty code in which faulty code that was seen at output of tolerant gates will be corrected at following tolerant gates by acceptance codes of Class_0 as Pole Code_0 and codes of Class_1 code as Pole Code_1 codes.

According this process, correction mechanism requires a following tolerant gate to mask errors resulting from previous tolerant gates. This mechanism leads errors at the last gates of the system stay uncorrected. Unfortunately, the last gates of system will be devoid of correction and they are vulnerable components of system in term of fault tolerance. For instance, in the Figure 7, faults appearing on T_Not gates connected to Q and Q̄ ports will not be correctable by tolerant coding mechanism.

## 4. Simulation Strategy And Some Results:

Faults on gates were reduced to an error bit on the wire connected to output of faulty logic gates. Error bit on a wire is implemented by complement of its correct value as seen Figure 12 in simulation.

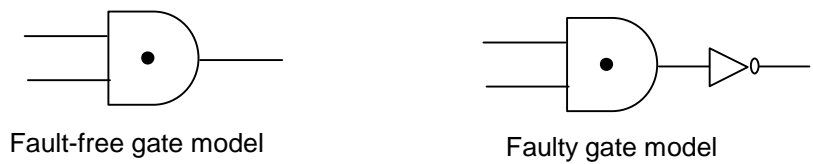

Fault-free gate model    Faulty gate model

**Figure 12.** Faulty *And* gate model made of fault-free And gate and an inverter for fault insertion

A fault on a gate was assumed to result an error bit at the output of the gate, which would be expectedly complement of error-free output. Error bits satisfying error probabilities were inserted to lines, which is connected to outputs of faulty gates. Obtained results after 1000 bits processing for fault probability of 0.005 was demonstrated by error graphs in the Figure 14, 15 and 16. In error graph, vertical axis represents processed input data up to 1000 and horizontal axis represents nets in EXOR with $(2,5)_3$ coding seen in Figure 13. Every line in the line graph represents error seen on the three bit nets. Light blue lines represent one-bit error on the net, yellow line represents two errors bit on the net and red line represent the case that all three bit of net in error. In the simulation, we used tolerant gates designed by truth table. In the Figure 14, we applied error insertion to tolerant gate connected to net3. As seen in the figure, all one-bit errors (Light blue lines) were corrected but two bit errors (Yellow lines) lived and reached to net7, which is output of EXOR.

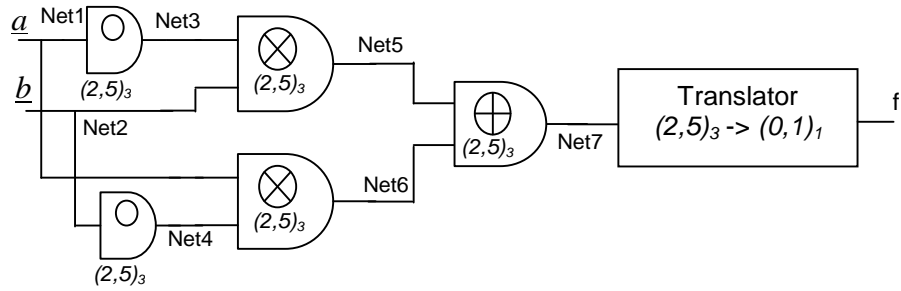

**Figure 13.** EXOR function implemented by $(2,5)_3$ coding to be used in simulations

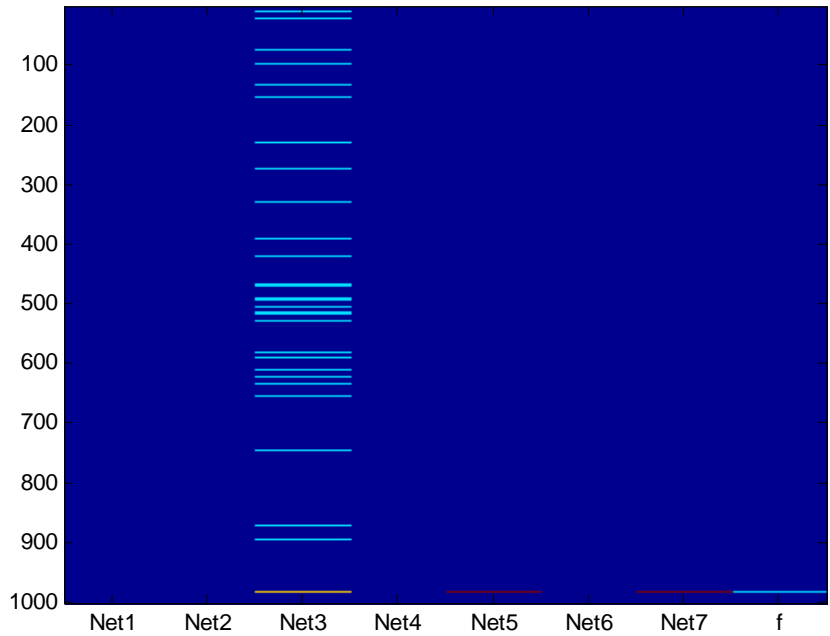

**Figure 14.** Error insertion to tolerant gates component connected to net3

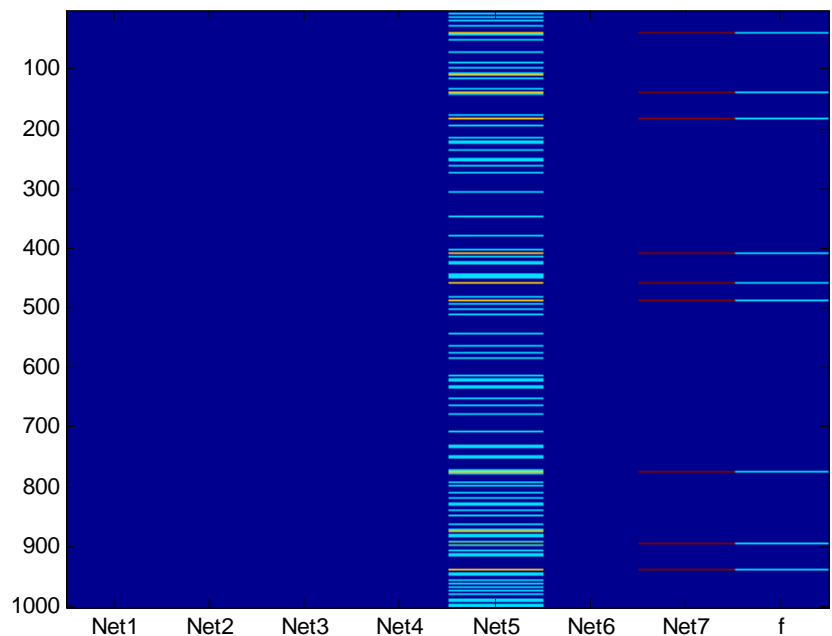

**Figure 15.** Error insertion to tolerant gates component connected to net5

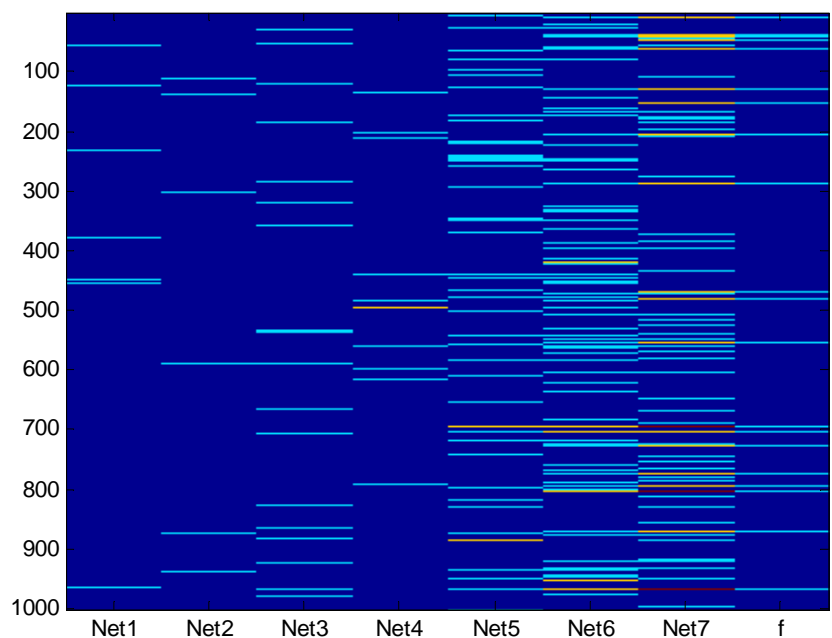

**Figure 16.** Error insertion to whole system

In the Figure 15, we applied error with higher probability on the tolerant gate connected to net5. As seen, only yellow and red error lived and reached to output. Light blue errors were corrected by $(2,5)_3$ coding by following tolerant gates. In the Figure 16, we applied error with probability of 0.005 to whole system except Translator. A few of error could be reached to fault-free translator output $f$.

In order to better compare $(2,5)_3$ tolerant coding and $(0,1)_1$ conventional coding performance, availability and tolerance rate were compared in the Figure 17 and 18. Availability was defined as rate of correct results to total results and expressed as following,

$$A = \frac{Cr}{Tr} \qquad (24)$$

,where $Cr$ is number of correct output and $Tr$ is number of total output. Tolerance rate ($To$) was defined as rate of incorrect results ($Icr$) to total number of error bits ($Ter$) inserted to all nets.

$$To = \frac{Icr}{Ter} \qquad (25)$$

For the simulation results seen in the Figure 17 and 18, error probability of logic gates is increased from 0.01 to 0.2 by 0.01 steps. EXOR circuit with $(2,5)_3$ fault tolerant coding was seen to superior fault tolerance and availability performance under uniform distributed random error insertion.

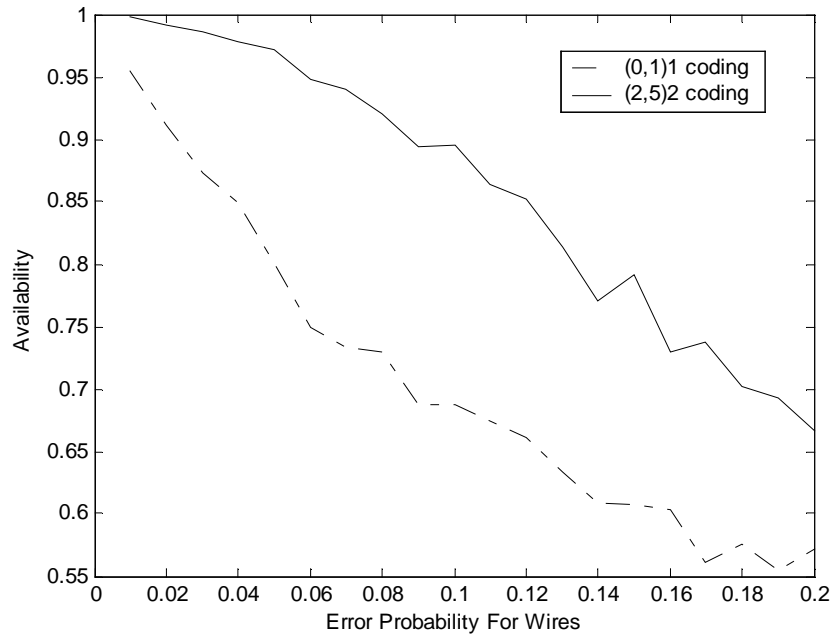

**Figure 17.** Availability of the $(2,5)_3$ tolerant coding and $(0,1)_1$ conventional coding

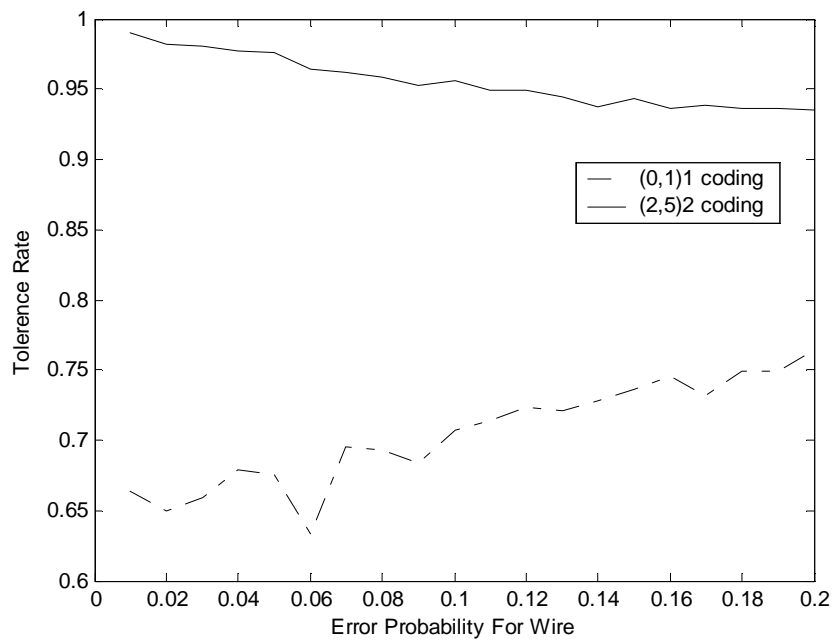

**Figure 18.** Tolerance rates of the $(2,5)_3$ tolerant coding and $(0,1)_1$ conventional coding

In the Figure 19, TMR supported EXOR function implemented by $(0,1)_1$ conventional coding was compared with $(2,5)_3$ tolerant coding in availability. In the simulation, TMR circuitry supporting 3 modules of EXOR logic was assumed as fault-free. It is seen from Figure 19 that $(2,5)_3$ tolerant coding has better availability than conventional TMR.

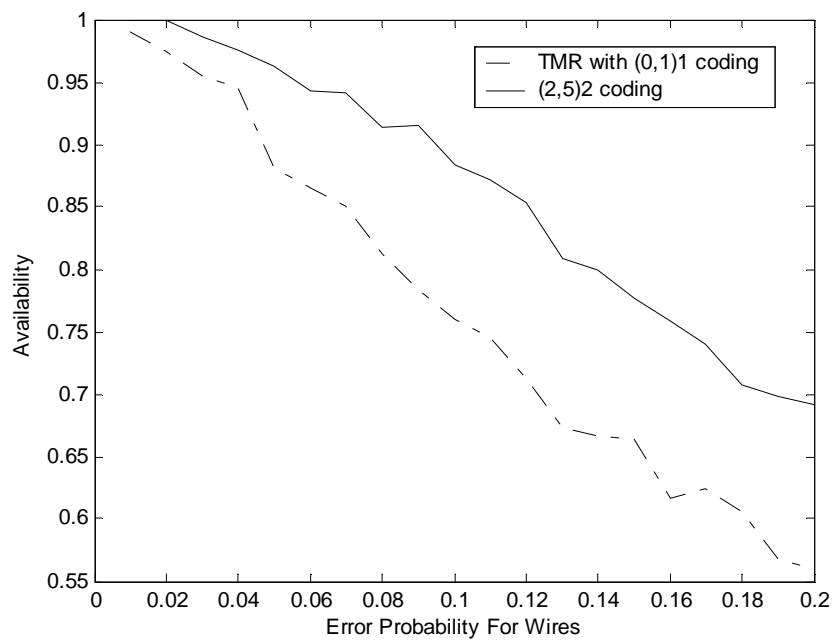

**Figure 19.** Availability of the $(2,5)_3$ tolerant coding and TMR supported $(0,1)_1$ conventional coding

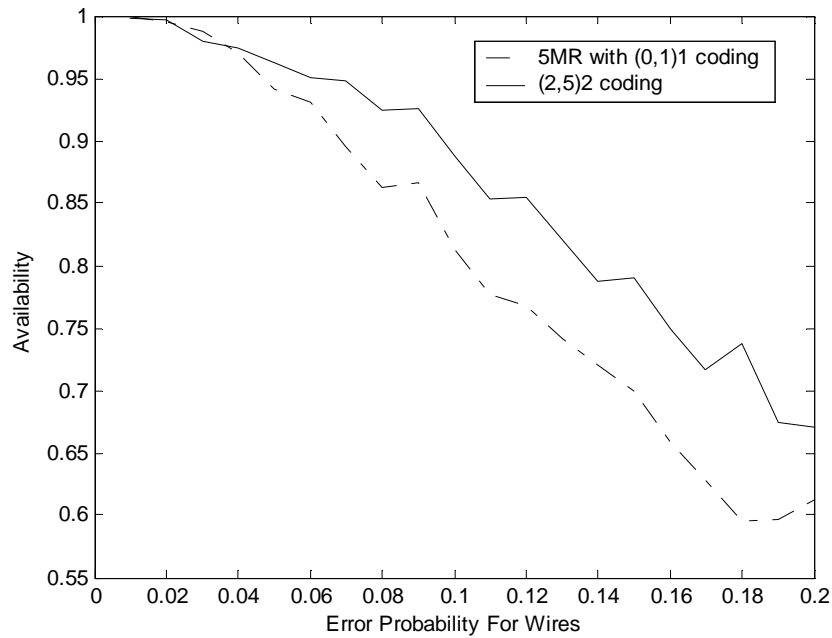

**Figure 20.** Availability of the $(2,5)_3$ tolerant coding and 5MR supported $(0,1)_1$ conventional coding

It is seen from Figure 20 that $(2,5)_3$ tolerant coding has slightly better availability than conventional 5MR.

## 5. A Way For Developing Information Processing System By Components With High Error Probability:

Advancing information processing systems such as nano-systems, quantum information processing structure or bio-information processing systems, basic components of such technology are not enough robust against environmental noise and structural defects. These components have high error probability in their operations. In order to develop more reliable information processing system made of such components, fault tolerance techniques should be invented to deal with errors. Fault tolerant coding technique introduced in this paper may contribute these types of applications having low reliability. Enough tolerant coding in higher bits hamming spaces may be discovered to implement practically reliable systems. In such technologies, component can yield more level than two. Two of these level can assigned for pole codes and the rest levels in output span can be grouped in Class_0 or Clas_1 according distance to poles in valid metric of systems. For some systems, valid metric for distance definition may be probability distributions or any depending on other statistical or physical parameters. Pole codes and transition mechanisms described in the paper gives an aspect researchers working on such new technology about how to handle output span of their components in order to be implement logic functions in a fault tolerant manner. In Figure 21, for probabilistic models, tolerant coding methodology is applied in similar manner discussed for logic system.

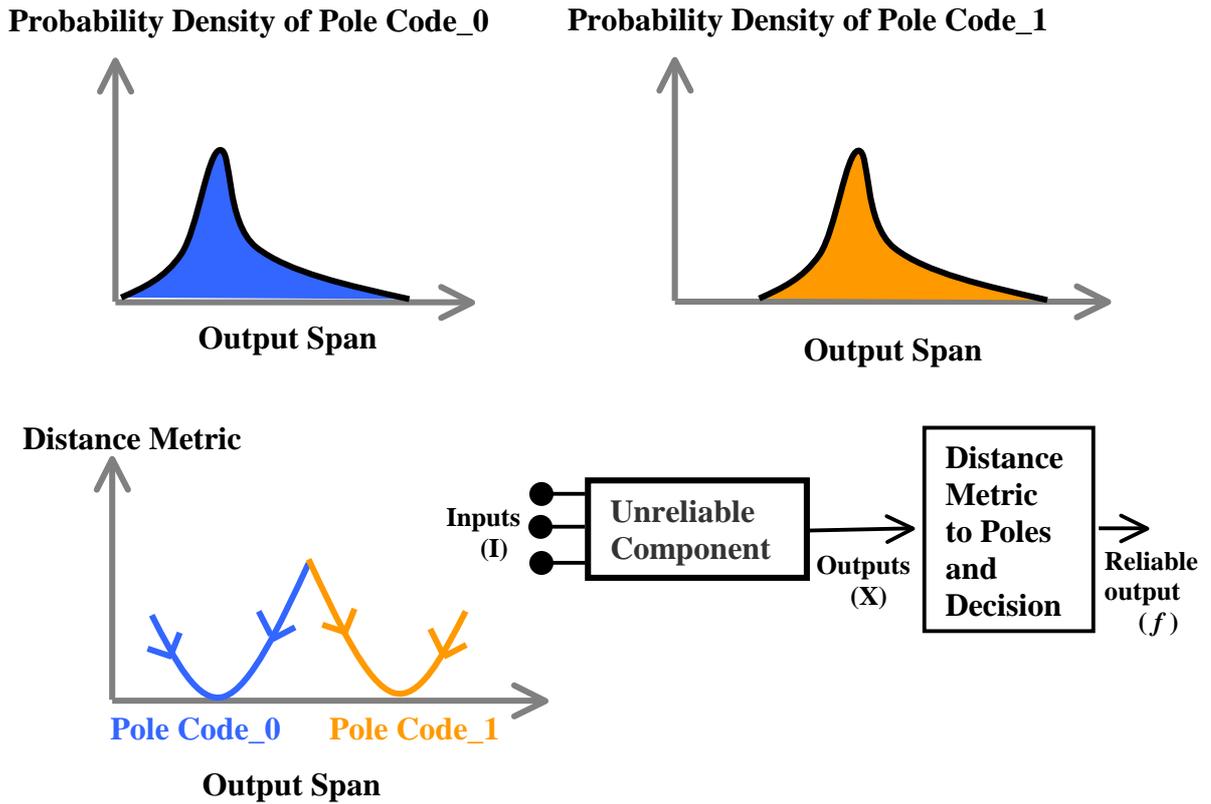

**Figure 21.** Tolerant coding representation in probability space

For an unreliable component, which has high error rates at output $X$, we define a distance metric on the span of output $X$ such that it complies with component functionality. Lets select pole codes in output set $X$ as $x_0$ and $x_1$. For a convenient distance metric ($\|x_i, x_j\|$) and pole codes ($x_0, x_1$) selection, following condition should be satisfied,

1- $\|x_0, x_1\|$ should be maximum in the space. (For selected distance metric and pole codes, distance between pole codes $x_0$ and $x_1$ should be maximum in space)

2- Class_H must be empty set. All elements of $X$ other than poles $x_0$ and $x_1$ should be reside in Class_0 or Class_1. This condition prevents '*unknown result state*', which is undesired state for Boolean based information processing systems.

If one finds a convenient distance metric ($\|x_i, x_j\|$) and pole codes ($x_0, x_1$) complying these condition, $f$ reliable output applying tolerant coding can be written as,

$$f = \begin{cases} x_0 & \|x_i, x_0\| \geq \|x_i, x_1\| \\ x_1 & \|x_i, x_0\| < \|x_i, x_1\| \end{cases} \quad (26)$$

After discussion for promising new information processing technologies, let us turn back to today's digital design technology based on silicon-based technologies. In order to obtain higher error correction performance, lets us investigate higher hamming space.

In the 5 bits hamming space, it is possible to select pole codes with 5-bit distance and fully correctable faulty codes with up to two bits distant to poles. So, it can correct error bits up to two bits. For example, $(10,21)_5$, $(0,31)_5$ …etc. Lets survey $(10,21)_5$ tolerant coding structures;

$$PoleCode\_0 = 2, \ PoleCode\_1 = 5$$
$$Class\_0 = \{0,2,3,4,6,8,9,11,12,14,15,18,24,26,27,30\}$$
$$Class\_1 = \{1,5,7,13,16,17,19,20,22,23,25,28,29,31\}$$

Transition graph for $(10,21)_5$ tolerant coding is given in the Figure 22.

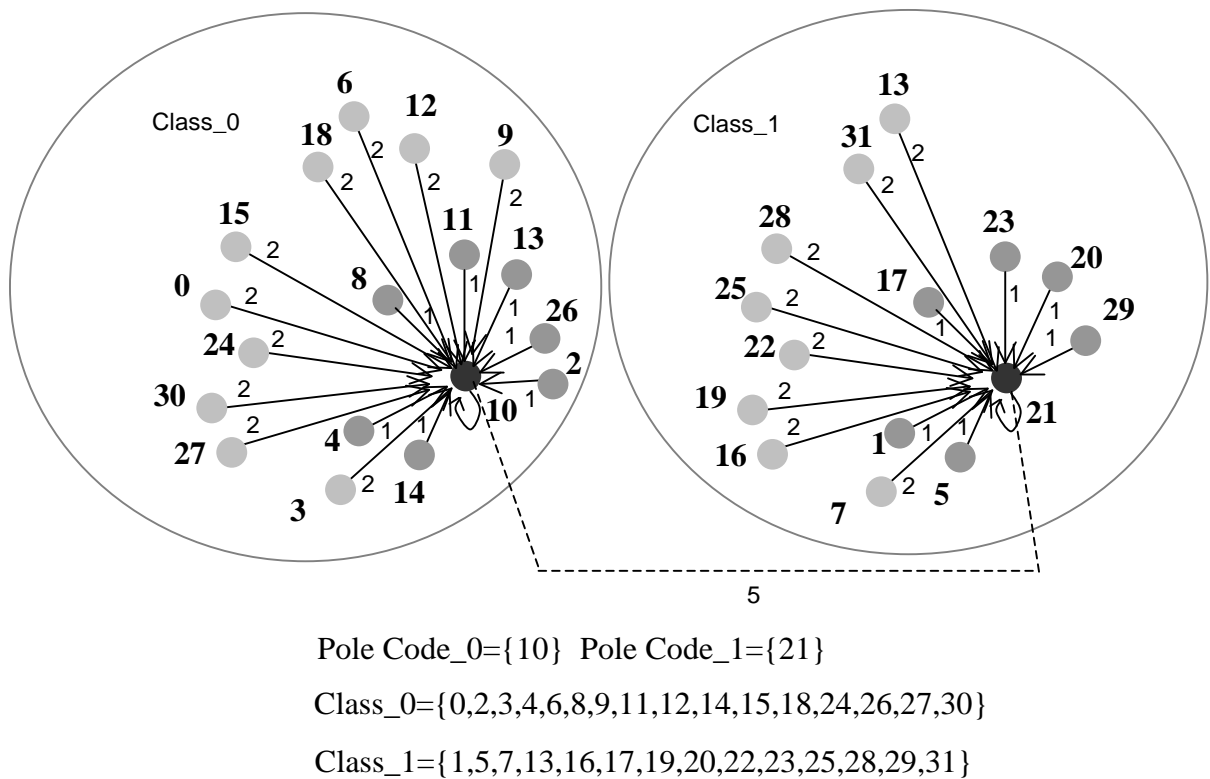

Pole Code_0={10}  Pole Code_1={21}

Class_0={0,2,3,4,6,8,9,11,12,14,15,18,24,26,27,30}

Class_1={1,5,7,13,16,17,19,20,22,23,25,28,29,31}

**Figure 22.** An example of transition graphs for a three-bits coding with three-bit distance ($(2,5)_3$ tolerant coding)

By general point of view, for *n*-bit hamming space, it would be possible tolerant coding that can produce pole codes with *n*-bit hamming distance and faulty code correction up to $n/2$ bits distance to pole codes. So, it may correct error bits up to $n/2$. One can find higher bit tolerant coding providing enough reliability to system composed of very noisy components.

An important point to remind is that designing in *n*-bit hamming space for a fault tolerant coding may consume large amount of resource and complexity may be very high. Consumed resource should be affordable for the obtained in availability.

## 6. Conclusions:

In this paper, fault tolerant coding, their properties and design concepts were briefly introduced for the hamming spaces up to three bits. Fault tolerant codes and their implementation by mean of conventional logic technologies were addressed in a systematic manner and a simplified and common design techniques that were applicable for all fault tolerant coding were developed. Expanding hamming space to higher bits gives us opportunity to assign some additional codes for fault detection and correction proposes in the logic system.

Fault tolerant coding highly increases the number of conventional logic gates and their interconnectivity in order to form higher hamming distances in hamming spaces. Transition mechanism acted substantial role on correction of error resulting from faults of the logic gates or wires.

For the future works, new optimum fault tolerant codes should be researched for discovering higher availability and lower consumption of resources.

## References:
[1] Richard W. Hamming. Error Detecting and Error Correcting Codes, Bell System Technical Journal 26(2),pp.147-160, (1950).
[2] Comtet, L. "Boolean Algebra Generated by a System of Subsets." in Advanced Combinatorics: The Art of Finite and Infinite Expansions, rev. enl. ed. Dordrecht, Netherlands: Reidel, pp. 185-189, (1974).
[3] Wil J.Van Gils, "A Triple Modular Redundancy Technique Providing Multiple-Bit Error Protection Without Using Extra Redundancy" IEEE Transactions On Computers, p.623, (1986)

## Acknowledgement:
This study was supported by ÖncüBilim System And Algorithm Laboratory for the memory of Serdar Onur Alagöz .